\documentclass[a4paper,12pt]{article}

\usepackage{cite}
\usepackage{graphicx}
\usepackage{amsmath}
\usepackage{setspace}
\usepackage{epstopdf}
\usepackage[lmargin=2.54cm,rmargin=2.54cm,tmargin=2.54cm,bmargin=2.54cm]{geometry}
\usepackage{appendix}
\usepackage{authblk}

\newcommand{\im}{\operatorname{Im}}

\newcommand{\Tr}{\operatorname{Tr}}

\title{Giant non-equilibrium vacuum friction: Role of singular evanescent wave resonances in moving media}
\author{Yu Guo}
\author{Zubin Jacob\thanks{zjacob@ualberta.ca}}
\affil{Department of Electrical and Computer Engineering, University of Alberta, Canada}
\date{\vspace{-5ex}}

\begin{document}
\maketitle
\onehalfspacing
\begin{abstract}
We recently reported on the existence of a singular resonance in moving media which arises due to perfect amplitude and phase balance of evanescent waves. We show here that the nonequilibrium vacuum friction (lateral Casimir-Lifshitz force) between moving plates separated by a finite gap is fundamentally dominated by this resonance.  Our result is robust to losses and dispersion as well as polarization mixing which occurs in the relativistic limit. 
\end{abstract}

\section{Introduction}
Nanoscale heat transfer beyond the black body limit has led to considerable interest due to simultaneous development of theoretical tools utilizing Rytov's heat transfer theory and experimental validation based on near-field measurements \cite{polder_theory_1971,pendry_radiative_1999,volokitin_radiative_2001,mulet_enhanced_2002,fu_nanoscale_2006,hu_near-field_2008,shen_surface_2009,rousseau_radiative_2009,biehs_nanoscale_2011,nefedov_giant_2011,jones_thermal_2012,guo_broadband_2012,guo_thermal_2013}. Surface waves play a key role in heat transfer which has been ascertained through bimetallic cantilever experiments\cite{shen_surface_2009} as well as near-field thermal emission spectroscopy\cite{de_wilde_thermal_2006,jones_thermal_2012}. This large radiative thermal energy transfer can be accompanied by momentum transfer and fluctuational forces which are nonequilibrium Casimir-Lifshitz forces\cite{teodorovich_contribution_1978,levitov_van_1989,pendry_shearing_1997,kardar_friction_1999,volokitin_near-field_2007,maghrebi_scattering_2013,guo_singular_2013,manjavacas_vacuum_2010,manjavacas_thermal_2010,maghrebi_spontaneous_2012}. 

One interesting case is that of Casimir plates at different temperatures in relative motion with a fixed gap between them\cite{pendry_shearing_1997,volokitin_theory_2008,philbin_casimir-lifshitz_2009,guo_singular_2013}. The heat transfer is accompanied by a lateral force opposing the motion (drag) since the exchanged photons carry preferential momentum along the direction of motion. This is fundamentally different from the stationary case where the symmetry of the configuration imposes the condition of net zero lateral momentum transfer.

In this paper, we outline a derivation of nonequilibrium vacuum friction utilizing the scattering matrix approach of heat transfer adapted to the case of moving media\cite{maghrebi_scattering_2013,guo_singular_2013}. Our aim is to study nanoscale light matter interaction in moving media which reveal subtle effects fundamentally distinct from conventional photonic media\cite{nezlin_negative-energy_1976,lambrecht_motion_1996,kardar_friction_1999}. In particular, we analyze the role of singular evanescent wave resonances on nonequilibrium vacuum friction\cite{guo_singular_2013}. We recently introduced a class of singular Fabry-Perot resonances in moving media which occurs due to a perfect phase and amplitude balance condition\cite{guo_singular_2013}. Such a combined phase and amplitude balance (PAB) condition can only occur for plates in relative motion and is fundamentally impossible in the stationary case. We considered before the role of this resonance for heat transfer\cite{guo_singular_2013} while here our focus is on the force accompanying the heat transfer (nonequilibrium vacuum friction).  We show the giant increase in the lateral drag force between the moving plates separated by a fixed gap. We trace the origin of this giant enhancement to the role of the unique phase-amplitude-balance (PAB) condition\cite{guo_singular_2013}. We also consider in detail the role of polarization mixing and show that the concept of the singular resonance is valid in the relativistic limit.

The paper is arranged as follows. In section 2, we briefly mention important earlier work in the field of vacuum friction and explain the origin of a singular resonance condition in moving media. In section 3, we derive a compact form for the scattering matrix of the moving plate including polarization mixing which occurs in the relativistic limit. Using this result we show the persistence of the singular resonance condition in spite of polarization mixing in section 4. Section 5 outlines a derivation of the nonequilibrium vacuum friction using the scattering matrix approach developed for heat transfer. Finally, in section 6, we analyze in detail the role of the singular resonance condition on the giant nonequilibrium drag force between moving plates. 

\section{Singular resonance in moving media}

Macroscopic van der Waals interactions and frictional forces proportional to the velocity goes back to the early work by Teodorovich who considered moving plates separated by a fixed gap\cite{teodorovich_contribution_1978}. The dissipative nature of this friction and a quantum field theoretic approach was outlined in Ref.~\cite{levitov_van_1989}. Pendry provided a derivation of this fluctuational drag force in the $T \rightarrow 0$ limit (quantum friction) making use of the zero point energy associated with the field fluctuations\cite{pendry_shearing_1997} wherein the macroscopic reflection coefficients play a key role . Other macroscopic approaches have also been recently developed\cite{silveirinha_exchange_2012,nesterenko_macroscopic_2014}. The stress tensor approach for friction between moving media closely following Lifshitz theory of Casimir forces was developed by Volokitin and co-workers\cite{volokitin_theory_2008}. Our approach closely follows that developed by Kardar et.al who interprets the emission and absorption of photons and resultant forces in terms of the near-field emissivity/absorptivity\cite{maghrebi_scattering_2013}. We strongly emphasize that the existence of this nonequilibrium vacuum friction between moving plates is now unanimously agreed to by everyone without any debate. 

\begin{figure} 
\centering
\includegraphics[scale=0.6]{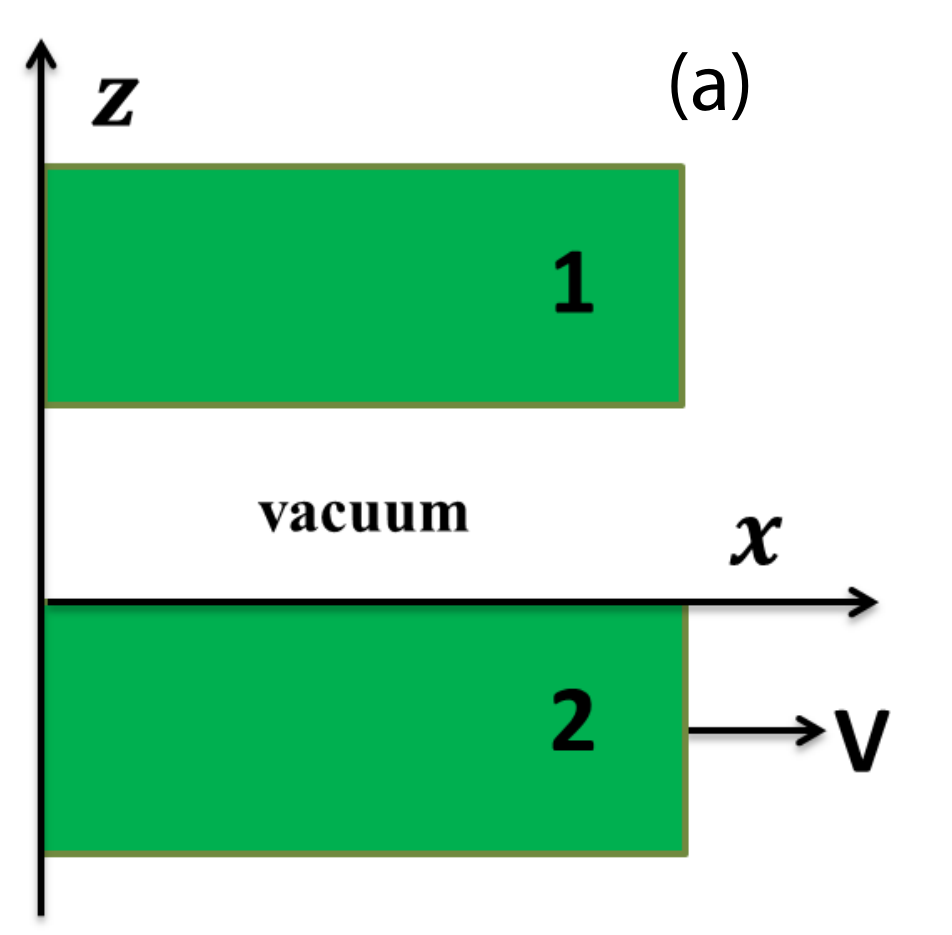}
\includegraphics[scale=0.5]{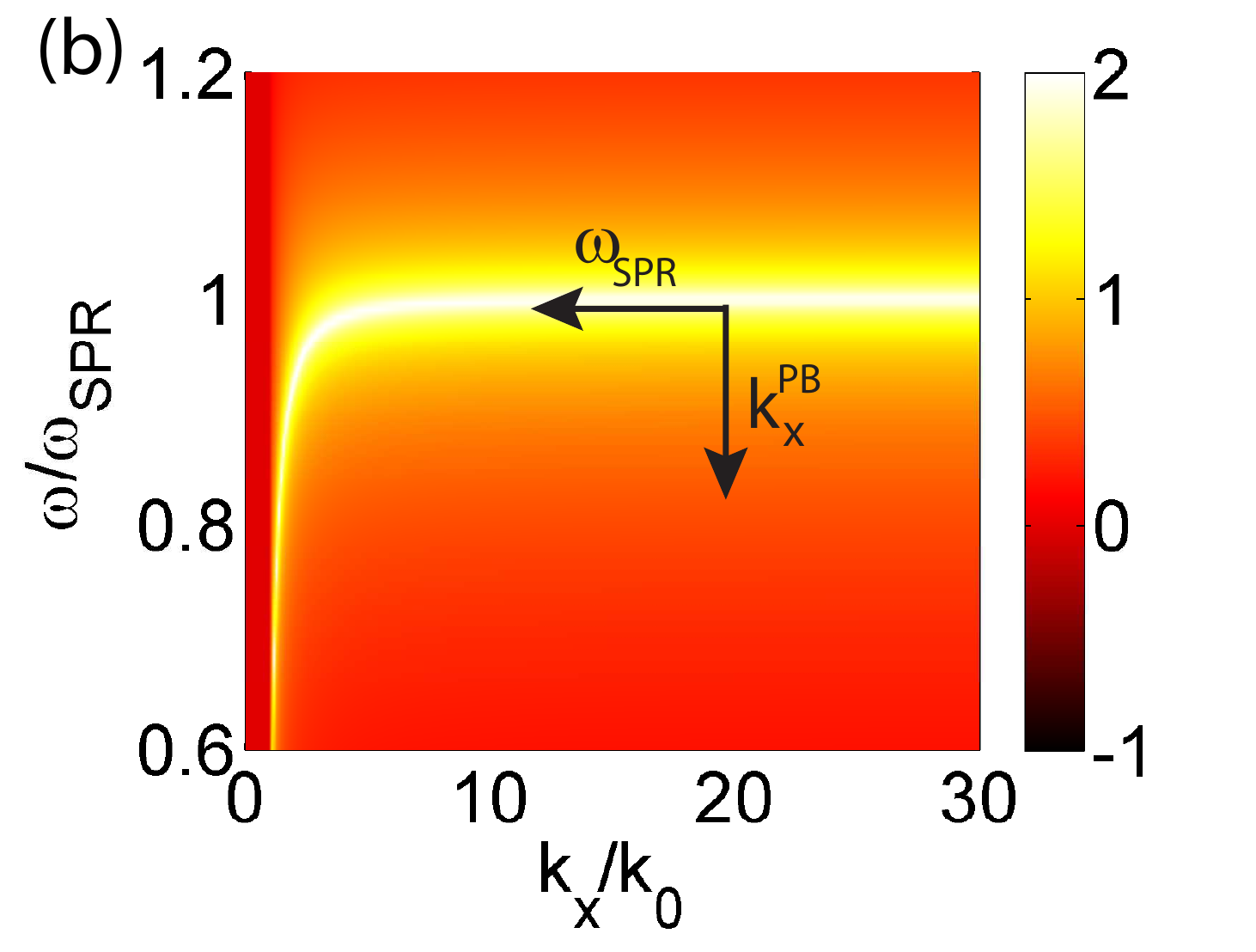}
\caption{\label{fig:schematic}(a) Schematic of the work. Two half-space plates composed of the same medium are separated by a vacuum gap. One plate is moving at a constant velocity V along the its interface (x direction). (b) Magnitude of reflection coefficient from a stationary plate of Drude metal in log scale, which  depicts the enhancement in the reflection coefficient due to the surface plasmon resonance (SPR). The special phase balance wavevector is $k_x^{PB} \approx 20k_0$ at a velocity of $V=c/10$ and $\omega_{SPR}= 10^{15}$Hz which leads to the unique mapping of the SPR resonance frequency of the moving plate to the negative SPR resonance frequency of the stationary plate ($\omega_{SPR} '=-\omega_{SPR} $).}
\end{figure}

We showed recently the existence of a unique phase-amplitude-balance (PAB) condition for surface plasmon polariton waves supported by moving metallic plates separated by a finite gap (Fig.~\ref{fig:schematic}(a))\cite{guo_singular_2013}. Consider a plane wave given by $e^{i(\vec{k}\cdot \vec{r}-\omega t)}$ incident from vacuum on an interface (x-y plane) moving at velocity $V$ parallel to the x-axis. The moving plate will then perceive a Doppler shifted frequency and wavevector \cite{kong_electromagnetic_1990}, 
$\omega '=\gamma (\omega -k_{x}V)$, $k_{x}^{'} =\gamma (k_{x} -\beta k_{0})$
where $\beta ={V/c}$, $\gamma ={1/ \sqrt{1-\beta ^{2}}}$.

The Doppler shift can lead to a unique condition when the frequencies in the moving and stationary frames are exactly the opposite of each other 
\begin{equation} 
\omega' =-\omega    
\end{equation} 
which occurs at a special wavevector (phase balance wavevector), 
\begin{equation}\label{pbk}
(k_{x}^{PB})^{'}=k_{x}^{PB} =(1+\frac{1}{\gamma})\frac{\omega}{V}.   
\end{equation}

At this specific wavevector, the conventional Fabry-Perot resonance condition 
\begin{equation}\label{FP}
\Delta_c=1-r_1r_2e^{2ik_zd}=0   
\end{equation}
takes the form 
\begin{equation}
\Delta_c=1-r_1(\omega)r_2(-\omega)e^{-2|k_z|d}=1-|r_1(\omega)|^2e^{-2|k_z|d}=0,   
\end{equation}
where $k_{z}$ is the wavevector along the perpendicular direction, $d$ is the gap between the plates, $r_1$ and $r_2$ are the reflection coefficients at the two interfaces, and we have used $r_2(-\omega)=r_1^*(\omega)$. 

If we function at the frequency where the metallic plates support a surface plasmon resonance (SPR)\cite{raether_surface_1988,maier_plasmonics:_2007}, we can achieve amplitude enhancement such that $|r(\omega_{SPR})|>1$. The perfect amplitude balance between evanescent wave enhancement at the plate and amplitude decay in the gap is achieved at
\begin{equation}
 d_0=\ln |r_1|/|k_z|
\end{equation} 
Thus one can fully satisfy the phase and amplitude balance (PAB) condition to achieve a singular resonance. The SPP dispersion of Fig.1(b) shows the existence of a surface wave resonance (field enhancement) with large wavevectors in the local limit. The waves with the specific phase balance wavevector (denoted by arrow) on the dispersion relation represent those SPP waves which are Doppler shifted exactly to their negative frequency counterpart. For these SPP waves, a unique Fabry-Perot resonance is achieved with $\Delta_c$ being identically zero which can never occur in stationary passive plates due to causality\cite{landau_electrodynamics_1984}.  We emphasize that this resonance relies on conversion of mechanical energy of motion to electromagnetic energy and gives a unique singularity in the Fabry-Perot multi-reflection factor ($1/\Delta_c$) that is routinely encountered in various phenomena. 


\section{Scattering matrix of a moving plate with polarization mixing}

Moving plates perceive Doppler shifted frequencies and hence the reflection coefficients of evanescent waves bouncing between them are fundamentally different from the textbook case of stationary Fabry-Perot plates.  Generally speaking even for moving isotropic media, the reflected waves of p-polarized incident waves will have s-polarized components, and vice versa. This is called polarization mixing due to motion. Here, we provide a succinct and generalized form of the scattering matrix of a moving plate and show the existence of the singular condition despite effects like polarization mixing that occur at relativistic velocities\cite{guo_singular_2013}. 
The scattering matrix of the moving plate $S$ can be expressed by 
\begin{equation} 
S =\left[\begin{array}{cc} {r_{ss} } & {r_{sp} } \\ {r_{ps} } & {r_{pp} } \end{array}\right].  
\end{equation} 

The  reflection coefficients are  
\begin{equation} 
r_{ss} =r_{s}^{'} a^2 -r_{p}^{'} b^2,   
\end{equation} 
\begin{equation} 
r_{sp} =-(r_{s}^{'}+r_{p}^{'} )ab,  
\end{equation} 
\begin{equation}  
r_{ps} =(r_{s}^{'}+r_{p}^{'} )ab =-r_{sp},  
\end{equation} 
\begin{equation} 
r_{pp} =r_{p}^{'}a^2 -r_{s}^{'}b^2,  
\end{equation} 
where $r_{s}^{'}$ and $r_{p}^{'}$ are the reflection coefficients in the co-moving frame for s- and p-polarized waves, respectively. And the factors 
\begin{equation} 
a={\gamma(k_{\rho }^{2} -\beta k_{0} k_{x} ) \Big / (k_{\rho }^{'} k_{\rho })}, 
\end{equation} 
\begin{equation} 
b={\gamma(\beta k_{y} k_{z}) \Big / (k_{\rho }^{'} k_{\rho })},
\end{equation} 
which obey
\begin{equation}
a^2+b^2=1. 
\end{equation}
Here $k_{\rho}=\sqrt{k_{x}^{2} +k_{y}^{2}}$, $k_{x}^{'} =\gamma (k_{x} -\beta k_{0} )$ and $k_{\rho}^{'} =\sqrt{(k_{x}^{'})^{2} +k_{y}^{2}}$. Note polarization mixing will not be a significant effect at non-relativistic velocities due to the factor $\beta$ in the off-diagonal elements of the reflection tensor ($r_{sp}$ and $r_{ps}$). 

It is interesting to note that along symmetry directions dictated by the direction of motion, polarization mixing disappears regardless of the velocity. From the above generalized scattering matrix, the case of $k_{y} =0$ simplifies to 
\begin{equation}
r_{ss} (\omega,k_{x})=r_{s}^{'} (\omega',k_{x}^{'}),  
\end{equation} 
\begin{equation} 
r_{pp}(\omega,k_{x})=r_{p}^{'} (\omega',k_{x}^{'}),  
\end{equation} 
and 
\begin{equation} 
r_{sp} =r_{ps} =0.  
\end{equation} 

The above equations have a simple but important interpretation. The off-diagonal components signifying polarization mixing are identically zero when  $k_{y} =0$. Furthermore and more importantly, the reflection from a moving plate can be expressed simply as the standard Fresnel reflection from a stationary plate with a Doppler shifted frequency and wavevector eg: for p-polarized waves with $k_{y} =0$, $r_{p}^{mov} (\omega,k_{x})=r_{p}^{'}(\omega',k_{x}^{'})$. Specifically, at the relativistic phase balance wavevector (Eq.~\ref{pbk}), we have the unique Doppler mapping 
\begin{equation} \label{new_wk} 
\omega '=-\omega, \: k_{x}^{'} =k_{x}. 
\end{equation} 
i.e. the frequency shifts sign and the wavevector remains the same, consistent with the invariance of the four dimensional momentum vector. Thus $r_{p}^{mov}(\omega,k_{x})=r_{p}^{'}(-\omega,k_{x})$. Note $r_{p}^{'}$ is reflection from a stationary plate in the co-moving frame so we have
\begin{equation}
r_{p}^{mov}(\omega ,k_{x})=(r_{p}^{'}(\omega ,k_{x}))^{*},
\end{equation}
which is the complex conjugate of the reflection coefficient from the stationary plate in the lab frame. We have thus recovered the central result that $r_{2} =r_{1}^{*}$ at the phase balance wavevector.

\section{Generalized resonance condition between moving plates}

We now formulate the resonance condition in terms of the scattering matrices ($S_1,S_2$) of the plates. In  matrix form, the generalized Fabry-Perot resonance condition is that the determinant of the matrix $1-S_{1}S_{2}e^{2ik_{z}d}$ should be zero, 
\begin{equation}\label{eq:delta}
\Delta=det(1-S_{1}S_{2}e^{2ik_{z}d})=a^2D_{ss} D_{pp} +b^2D_{sp} D_{ps}=0.
\end{equation}
Here $D_{ss} =1-e^{2ik_{z}d} r_{1s} r_{2s}^{'}$, $D_{pp}=1-e^{2ik_{z}d} r_{1p} r_{2p}^{'}$, $D_{sp} =1+e^{2ik_{z} d} r_{1s} r_{2p}^{'}$, $D_{ps}=1+e^{2ik_{z}d} r_{1p} r_{2s}^{'}$. $r_{1(s,p)}$ are the reflection coefficients from the stationary plate, $r_{2(s,p)}^{'}$ are the reflection coefficients in the co-moving frame. Note the phase balance wavevector achieves the unique Doppler mapping $\omega '=-\omega, \: k_{x}^{'} =k_{x}$ and hence the various components of $\Delta$ have to be analyzed for this specific case. Due to the reality of fields, the reflection coefficients in the co-moving frame at frequency $-\omega $ are the complex conjugates of corresponding reflection coefficients in the lab frame at frequency $\omega$, $r_{2(s,p)}^{'}(-\omega,k_x,k_y)=r_{1(s,p)}^{*}(\omega,k_x,k_y)$. 

We now note the critical fact that at the phase balance wavevector we have
\begin{equation}
\im(D_{ss})=\im(D_{pp})=\im(D_{sp}D_{ps})=0.
\end{equation}
Thus the multi-reflection factor $\Delta$ which includes polarization mixing and relativistic effects is real valued at the phase balance vector. Furthermore, in the presence of surface waves there will always exist a critical distance when this multi-reflection factor is identically zero ($\Delta=0$). The main contribution of our work is this delicate phase balance and amplitude balance condition that has not been pointed out before.


The multi-reflection factor $\Delta$ exhibits a weak dependence on $k_y$ at non-relativistic velocities. Furthermore, waves with non-zero $k_y$ have enhanced damping compared to the case of $k_y=0$. Therfore $\Delta$ will reach its local minimum at $k_y=0$ where the damping factor $e^{-2|k_z|d}$ is smaller compared to the case with nonzero $k_y$. We now have a simplified form for the multi-reflection factor which is $\Delta=D_{ss}D_{pp}$. The s-polarized reflection coefficients exhibit no enhancement due to lack of surface waves ($|r_s|<1$), implying a positive $D_{ss}$, so to achieve a resonance i.e. $\Delta=0$, we arrive at the condition
\begin{equation}
D_{pp}(k_{x}^{PB},k_y=0)=1-\left|r_{p}(\omega)\right|^{2} e^{-2|k_z|d}=0. 
\end{equation}

We can adjust the distance $d$ to make this factor zero as long as there is surface wave enhancement for the phase balance wavevector ($\left|r_{p}\right|>1$).  Note that $|r_{p}|$ reaches its maximum at the surface plasmon resonance (SPR) due to evanescent wave enhancement ($\omega=\omega_{SPR}$). Thus the condition that $D_{pp}$ equals zero at this resonance frequency leads to the critical distance 
\begin{equation} \label{eq:d0}
d_{0}=\frac{V}{\omega_{SPR}}\ln \left|r_{p}(\omega_{SPR})\right| \sqrt{\frac{\gamma^2}{2+2\gamma}},   
\end{equation} 
For most cases of interest, we have $\gamma=1$, so $d_{0}=V\ln |r_{p}(\omega_{SPR})|/(2\omega_{SPR})$. Note that we achieve an upper bound on the critical distance. 
The role of deviations from the Drude model and limitations of amplitude enhancement due to non-locality has been analyzed in \cite{guo_singular_2013}. The analysis shows that the minimum velocity to observe the effect is of the order of the Fermi velocity of the electrons in the metal.  

\section{Derivation of non-equilibrium vacuum friction}
We now consider the effect of such a resonance on physical observables. We choose to study the momentum transfer between moving plates that leads to a drag force opposing motion: non-equilibrium vacuum friction. We emphasize that the non-integrable nature of 
this resonance can play an important role in various phenomena. Our derivation utilizes the scattering matrix theory of heat transfer\cite{maghrebi_scattering_2013,bimonte_scattering_2009} generalized to the case of moving media to recover the results known from the Maxwell's stress-tensor approach for Casimir forces\cite{volokitin_theory_2008,philbin_casimir-lifshitz_2009}. 
The net number of photons exchanged between the two plates is
\begin{equation} 
N=\Tr \left[(1-S_{2}^{\dag } S_{2} )D(1-S_{1} S_{1}^{\dag } )D^{\dag } \right](n(\omega,T_1)-n(\omega',T_{2})) 
\end{equation} 
for propagating waves (PWs), and 
\begin{equation}  
N=\Tr \left[(S_{2} -S_{2}^{\dag } )D(S_{1}^{\dag } -S_{1} )D^{\dag } \right](n(\omega,T_1)-n(\omega',T_{2} )) 
\end{equation} 
for evanescent waves (EWs), where $D={e^{ik_{z}d}}/({1-S_{1}S_{2}e^{2ik_{z}d}})$, $n(\omega,T)=1/(e^{\hbar \omega/k_BT}-1)$ is the Bose-Einstein occupation number. 

With the help of the scattering matrix and after some algebra, one can derive the expression for number of photons exchanged, 
\begin{align} \label{pn}
& N(\omega ,k_{x} ,k_{y})=(n(\omega ,T_{1})-n(\omega ',T_{2}))\bigg\{ \frac{1}{|\Delta|^{2}}e^{-2\im(k_{z} )d} \nonumber \\
& \Big[a^2(1-|r_{1p} |^{2} )(1-|r_{2p}^{'}|^{2})|D_{ss}|^{2}+b^2(1-|r_{1p} |^{2})(1-|r_{2s}^{'}|^{2} )|D_{sp} |^{2} +(p\to s)\Big] \bigg\},\: \mbox{PWs} \nonumber \\ 
& N(\omega,k_{x} ,k_{y})=(n(\omega ,T_{1})-n(\omega',T_{2}))\bigg\{ \frac{4}{|\Delta|^{2}}e^{-2\im(k_{z} )d} \nonumber \\
& \Big[a^2\im(r_{1p} )\im(r_{2p}^{'})|D_{ss}|^{2}-b^2 \im(r_{1p} )\im(r_{2s}^{'} )\left|D_{sp} \right|^{2} +(p \to s)\Big] \bigg\},\: \mbox{EWs}
\end{align} 
The symbol $p\leftrightarrow s$ denotes the terms that can be gained by permuting the indexes p and s of preceding terms, and the definitions of other symbols can be found in Eq.~\ref{eq:delta}. 
The dispersive force, i.e., the momentum transfer between the two plates\cite{maghrebi_scattering_2013}, is the product of the total number of exchanged photons and the momentum of a single photon $\hbar k_{x}$ ($\hbar$ is the Planck constant divided by $2\pi$)
\begin{equation} \label{force} 
f_{x} (\omega,k_{x},k_{y})=\hbar k_{x} N(\omega,k_{x},k_{y}).   
\end{equation} 
We also note that the energy transfer between the two plates is the product of the total number of photons exchanged and the energy of a single photon $\hbar \omega$.

The net dispersive force can be achieved by integrating all possible partial waves $\omega$, $k_{x}$ and $k_{y}$\cite{maghrebi_scattering_2013} in  the above Eq.~(\ref{force}). Note that the frequency $\omega$ should be positive. The friction can be calculated by 
\begin{equation} \label{total_force} 
F_{x} =\int _{0}^{\infty }\frac{d\omega}{2\pi } \int _{-\infty }^{\infty }\frac{dk_{x} }{2\pi} \int _{-\infty}^{\infty }\frac{dk_{y}}{2\pi} \hbar k_{x} N(\omega,k_{x},k_{y}).  
\end{equation} 
Note that $N$ has different expressions for propagating and evanescent waves (see Eq.~(\ref{pn})). We can then recover the results in Ref.~\cite{volokitin_theory_2008} which has a detailed calculation based on the stress tensor approach. 

\section{Results and discussion}
\begin{figure}
\centering
\includegraphics[scale=0.5]{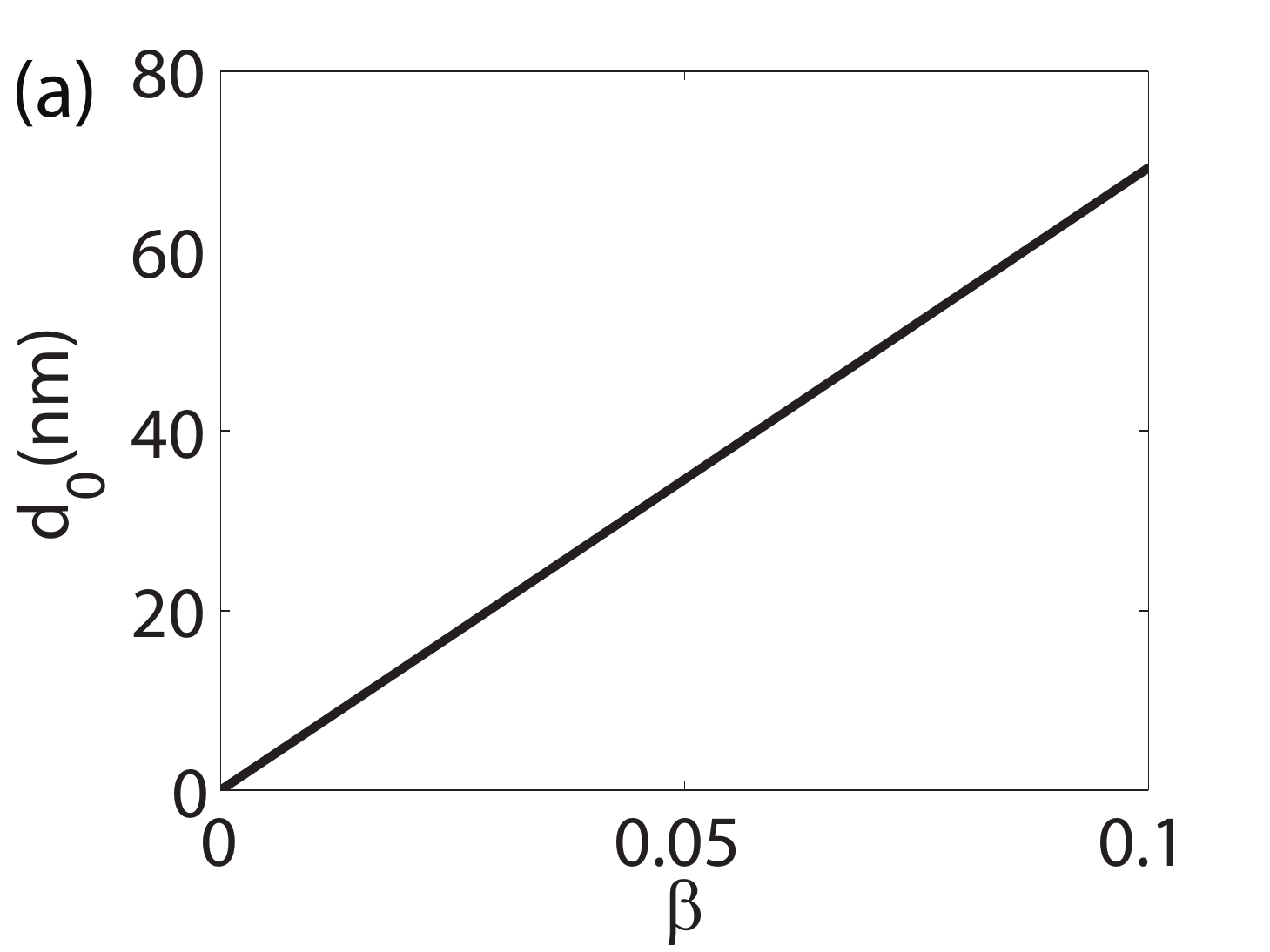}
\includegraphics[scale=0.5]{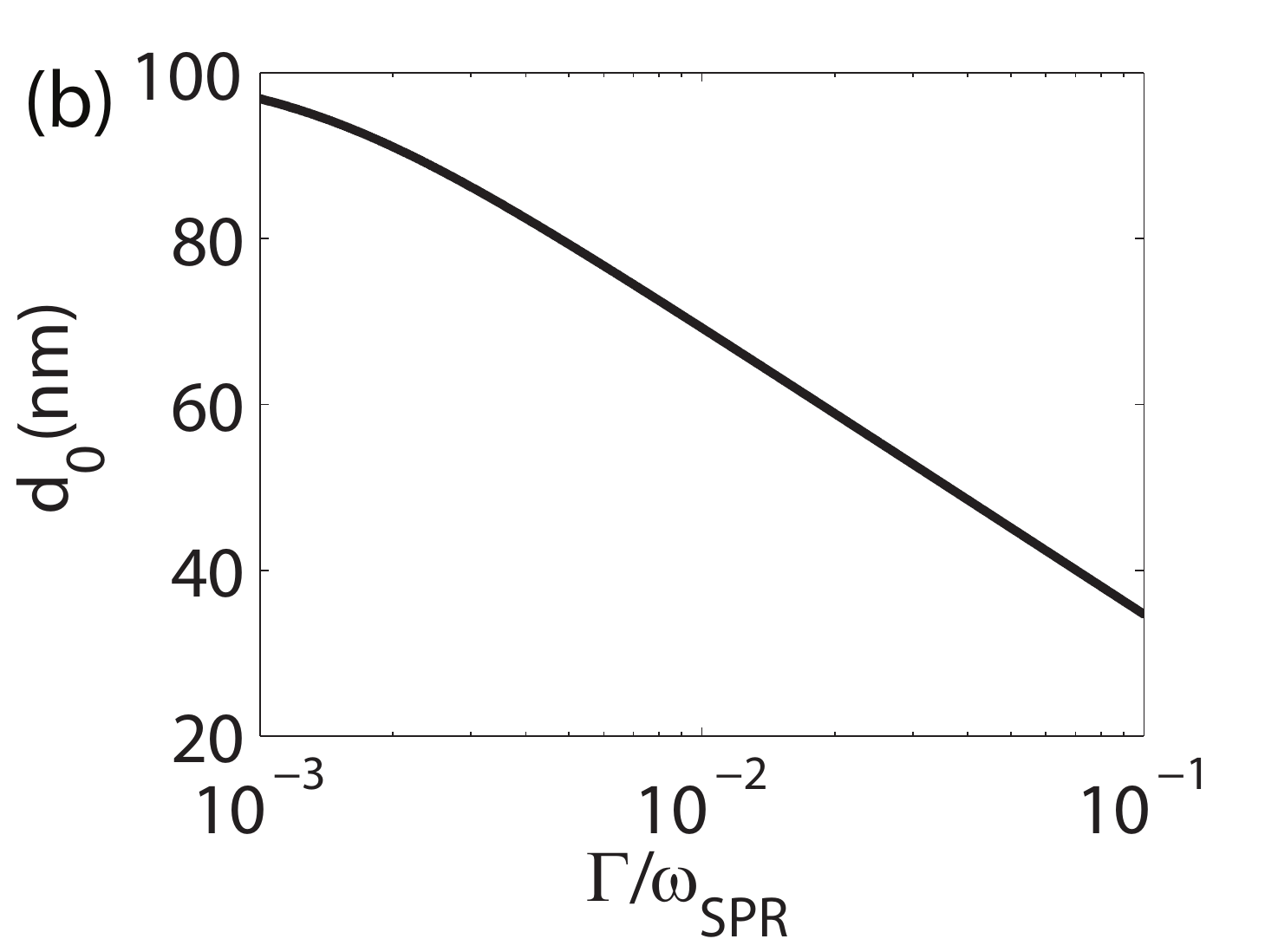}
\caption{\label{fig:d0} The opposing effects of increasing velocity and damping on the singular resonance condition. (a) Critical distance varying with moving velocity. The critical distance increases linearly versus moving velocity, in agreement with Eq.~\ref{eq:d0}. (b) As the Drude damping increases, the critical distance where the resonance occurs decreases. This is expected since the amplitude balance condition is sensitive to loss in the surface plasmon resonance and occurs only if the plates are in the extreme near-field.}
\end{figure}
Here we consider a Drude metal with frequency dependent permittivity given by $\epsilon(\omega)=1-2{\omega _{SPR}^{2} / (\omega ^{2} +i\Gamma \omega)}$ with the surface plasmon resonance (SPR) frequency $\omega _{SPR} =1\times 10^{15} {\rm Hz}$ and $\Gamma =0.01\omega_{SPR}$. The temperatures are chosen to be ${T}_{1}$=320K and ${T}_{2}$=300K. Note the resonant effect we are considering depends solely on the classical electromagnetic scattering matrix and the effect does not depend on the temperature difference. At the velocity of $V=c/10$, the critical distance $d_{0}$ that satisfies the singular resonance condition is about 70nm. In Fig.~\ref{fig:schematic}(b), we show the enhancement of reflection coefficients due to SPR and indicate the phase balance wavevector at this velocity. 

In Fig.~\ref{fig:d0}(a), we show the effect of moving velocity on the critical distance. It is clear that the critical distance increases linearly as moving velocity increases. We also examine the dependence of critical distance on the loss parameter $\Gamma$ in the Drude model, which affects the reflection coefficient, or the enhancement of evanescent waves. At the quasi-static approximation, we have $|r_{p}(\omega_{SPR})| = \omega_{SPR}/\Gamma$, so from Eq.~\ref{eq:d0} the critical distance $d_0$ decreases logarithmically as the loss parameter increases. In Fig.~\ref{fig:d0}(b), we observe that $d_0\sim -\log(\Gamma/\omega_{SPR})$, which is thus in agreement with our theoretic prediction. 

\begin{figure}
\centering
\includegraphics[scale=0.5]{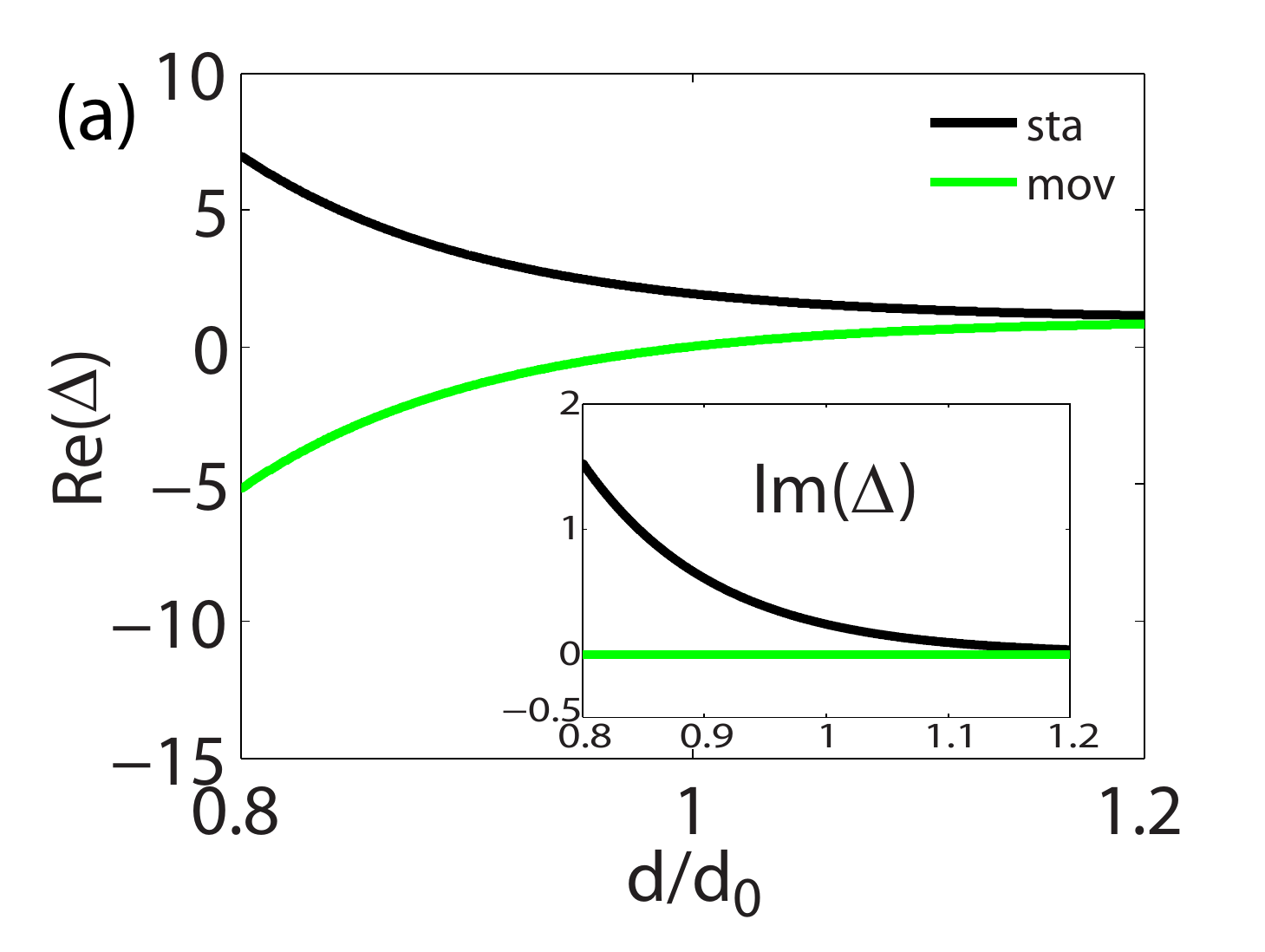}
\includegraphics[scale=0.5]{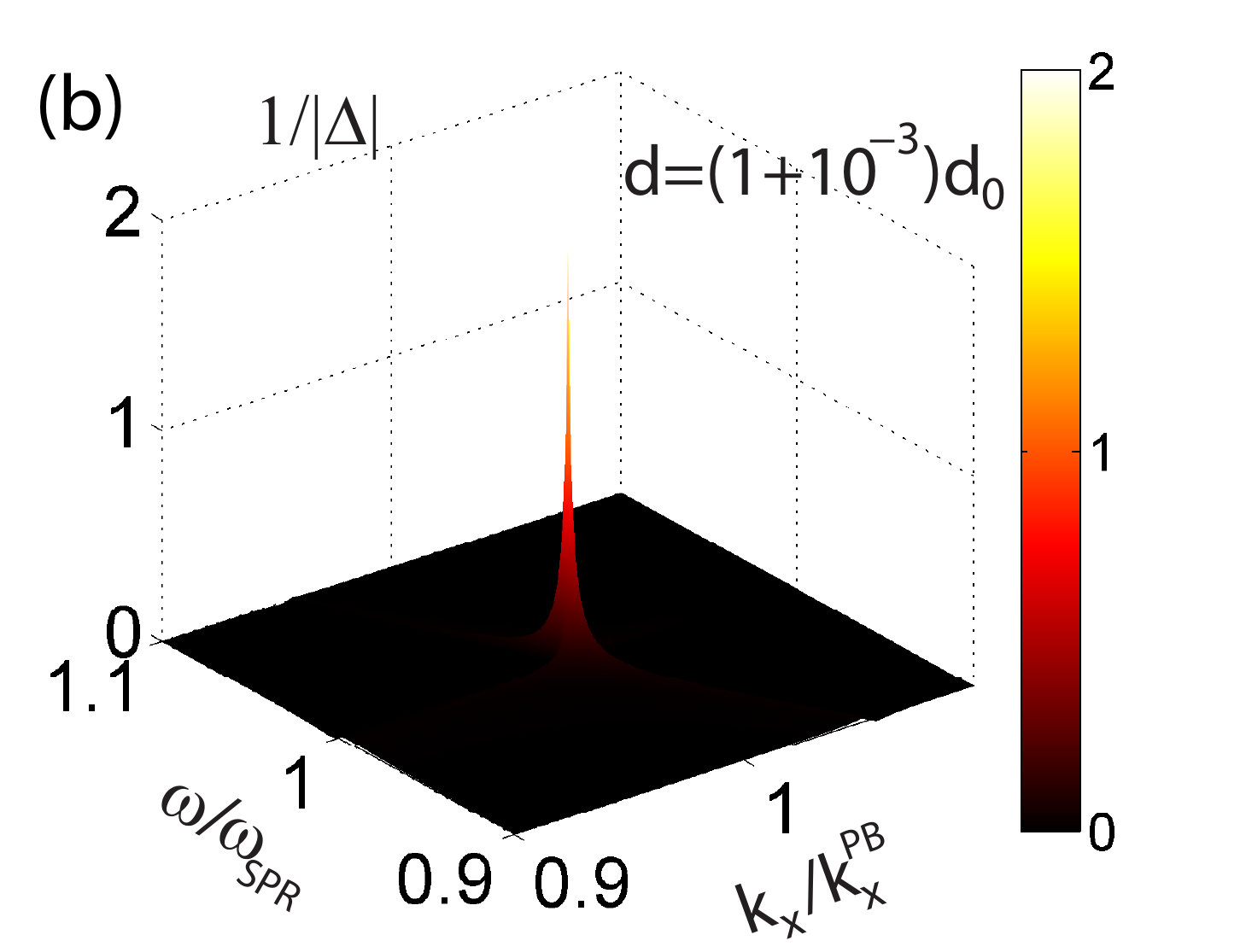}
\caption{\label{fig:delta} (a)Values of multi-reflection factor $\Delta$ as a function of distance at the phase balance wavevector and SPR frequency. The fact that $\im(\Delta)=0$ holds true for all distances in the moving plates, which is not valid for the stationary case. For moving plates at the phase balance wavevector, $\Delta$ crosses zero at the critical distance ($d_0$). However, for stationary plates, $\Delta$ can never be exactly zero, which is clear from the plot. (b)The denominator in the expression of photon transfer ($1/|\Delta|$) resolved by frequency and lateral wavevector $k_x$ at $d\to d_0^+$ ($d=(1+10^{-3})d_0$). Note the huge peak in the middle located at the phase balance wavevector and the SPR frequency, where the PAB condition leads to a vanishing $\Delta$. In both (a) and (b), we take $k_y=0$.}
\end{figure}

In Fig.~\ref{fig:delta}(a), we plot the multi-reflection factor $\Delta$ at various distances near the critical distance at the phase balance wavevector and the SPR frequency. In the moving plates, we can clearly see that $\Delta$ is purely real at the phase balance wavevector, which is a sign of phase balance and in agreement with our theory. Furthermore, $\Delta$ crosses zero at the critical distance, where the amplitude balance is fulfilled. However, for the stationary case, $\Delta$ can never be exact zero, which is clear in the plot. In Fig.~\ref{fig:delta}(b), we plot the denominator ($1/|\Delta|$) in the expression of photon transfer by varying frequencies and lateral wavevector ($k_x$). The distance is chosen to be close to the critical distance. We clearly see a huge peak located at the phase balance wavevector and SPR frequency, since at such a distance, the multi-reflection factor ($\Delta$) is very close to zero indicated by Fig.~\ref{fig:delta}(a). Thus we expect a giant photon number transfer at the PAB condition, which can further lead to giant momentum transfer between the moving plates. 

\begin{figure}
\centering
\includegraphics[scale=0.5]{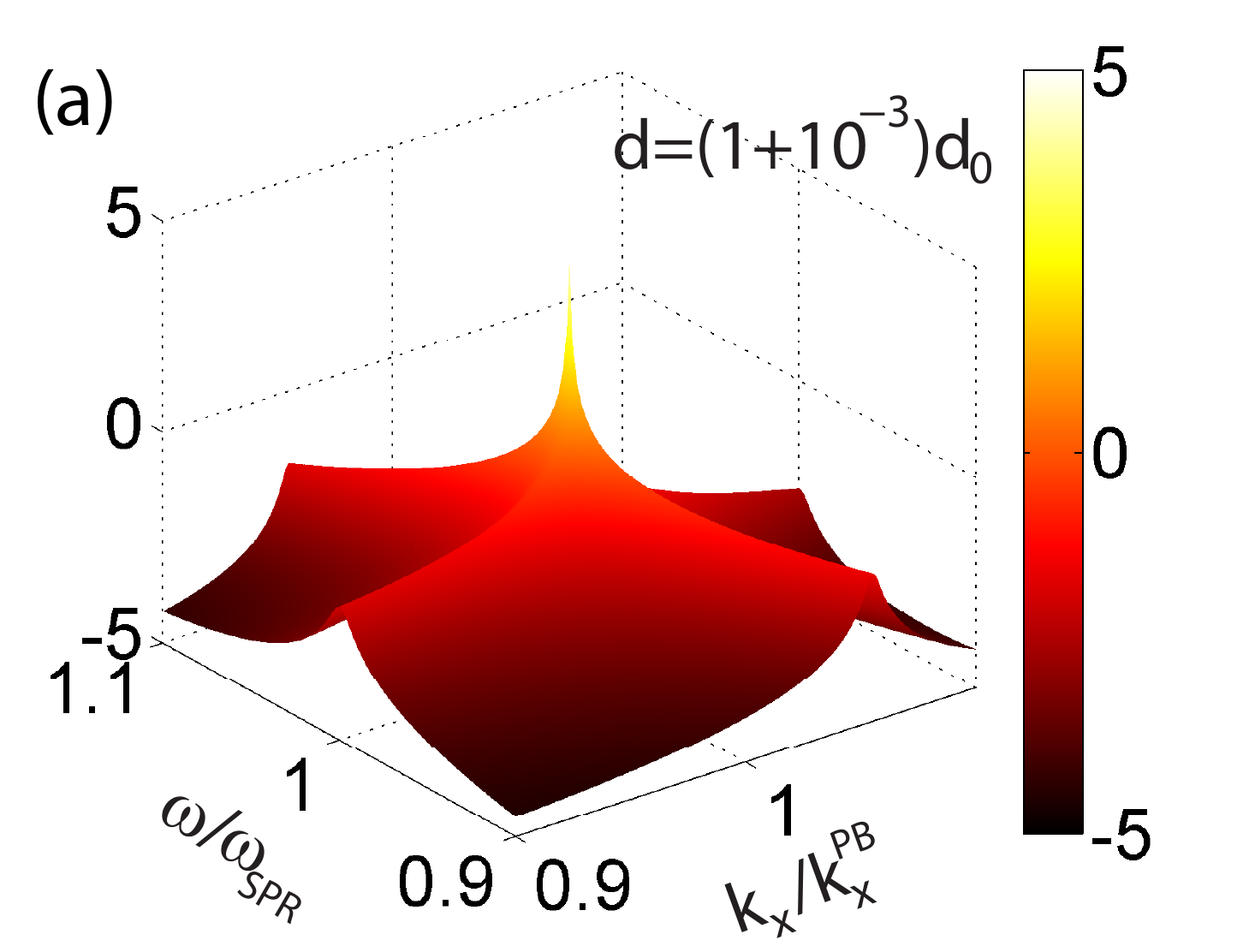}
\includegraphics[scale=0.5]{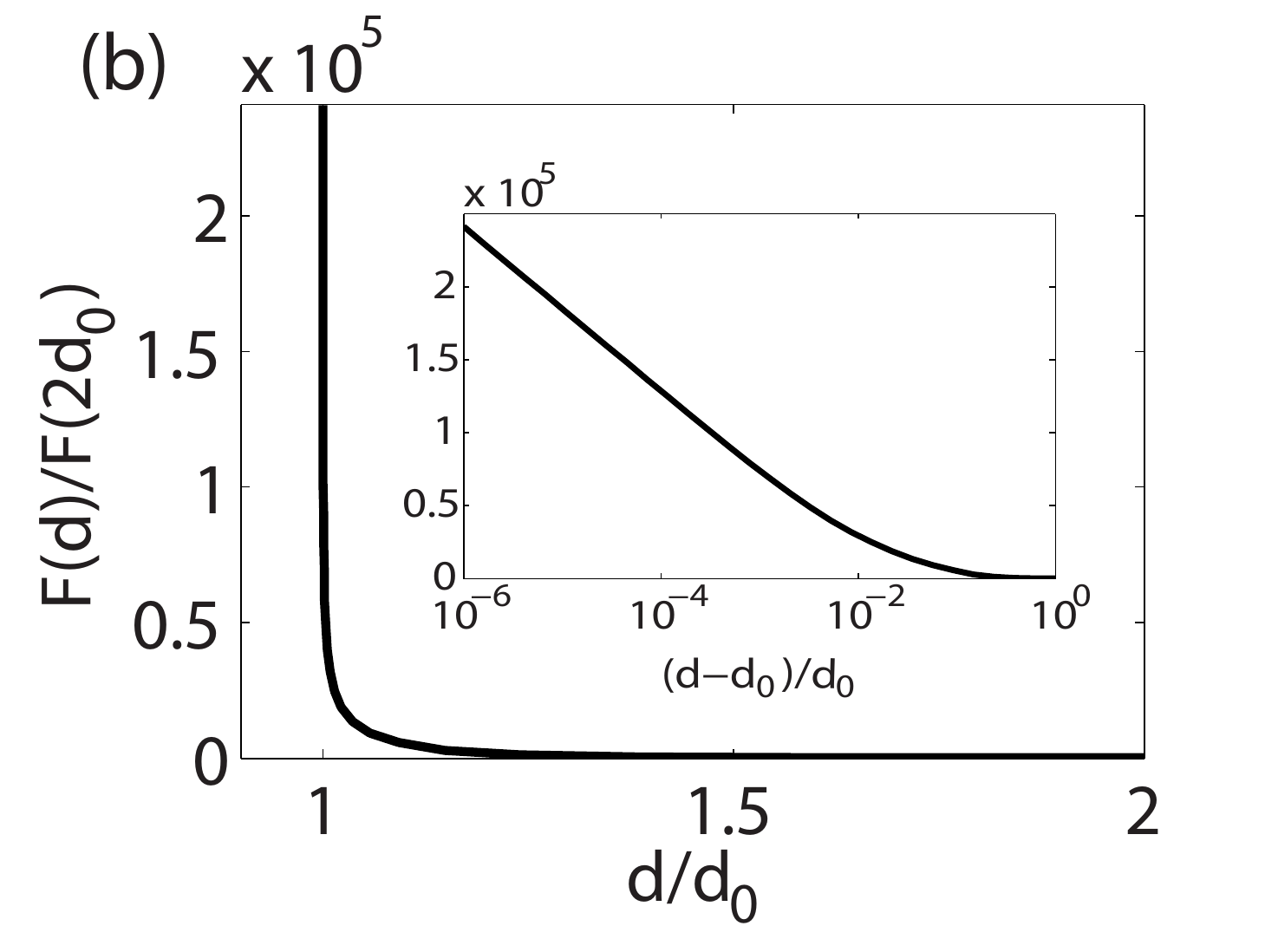}
\caption{\label{fig:force} (a) Contribution to momentum transfer resolved by frequency and lateral wavevector $k_x$ (normalized to the phase balance wavevector) at $d\to d_0^+$ ($d=(1+10^{-3})d_0$). The huge peak is due to the singular resonance that arises since the amplitude balance condition is satisfied when $d\to d_{0}^+$ and phase balance condition is satisfied at the phase balance wavevector. This leads to giant photon exchange and thus momentum transfer between moving plates at the singular resonance. (b)The distance dependence of friction at distances near ${d}_{0}$. The force increases dramatically near the critical distance $d_0$. In the inset, the x axis is in ($d/d_{0}-1$) and log scale. We clearly see a linear increasing behavior as d approaches d${}_{0}$. This is consistent with our theoretical scaling law which predicts a giant non-equilibrium vacuum friction.}
\end{figure}

In Fig.~\ref{fig:force}(a), we plot the spectrum of momentum exchanged according to their frequency and wavevector in the lab frame. 
Our result shows that as the plates are moved closer to the singular FP resonance condition ($d\to d_{0}^+$), a fundamentally new mechanism of photon exchange emerges. This is evident from Fig.~\ref{fig:force}(a) where photons with the phase balance wavevector and SPR frequency completely dominate the interaction. Note that this occurs when the frequencies in the co-moving frame and lab frame are equal and opposite, the condition for phase balance and also the enhancement of evanescent waves due to SPR compensate the decay of waves inside the gap, the condition of amplitude balance. Indeed, the multiple scattering term $\Delta$ becomes vanishingly small giving rise to a giant enhancement in the number of photons exchanged. 

In Fig.~\ref{fig:force}(b), the magnitudes of friction evaluated around the resonance at $d_{1} =2d_{0}$ and $d_{2} =(1+10^{-6})d_{0}$ are $2.34\times10^{-6}$N/m$^2$ and 0.563N/m${}^{2}$, respectively. We strongly emphasize the five order of magnitude increase in friction when the distance changes only by 70 nm. When the distance $d$ approaches the critical distance $d_{0}$, we predict the non-equilibrium friction F to scale with distance as\cite{guo_singular_2013}
$F\sim \ln \left[{d_{0}/(d-d_{0}})\right]$. We plot the friction vs. distance in Fig.~\ref{fig:force}(b) to verify the theoretical predictions. We clearly see that the force increases dramatically near the critical distance. We also see that the friction increases  as  $\ln \left[{d_{0}/(d-d_{0}})\right]$ when $d$ approaches $d_{0}$. 

We do not assume ideal mirrors \cite{kardar_friction_1999} and losses or dispersion are not an impediment to the singular resonance. At such a high velocity of $c/10$, our assumption of a local Drude model should be valid because the phase balance wavevector ($20k_0$) is not very large. However, for lower velocities, the corresponding phase balance wavevector can be significantly larger than free space wavevector, where the theory should be modified for electromagnetic interactions with large wavevectors\cite{mcmahon_nonlocal_2009,raza_unusual_2011}. Note, the only fundamental requirement is the enhancement in the reflection of coefficient of evanescent waves which is known to occur even in the presence of non-locality (eg: graphene plasmons\cite{koppens_graphene_2011,brar_highly_2013}). The role of the giant photon flux caused by the resonance on assumptions of macroscopic and local fluctuational electrodynamics will be analyzed in future work. 

\section{Conclusion}
In summary, we have provided a compact form of the scattering matrix of moving plates and derived the nonequilibrium vacuum friction formula through the scattering matrix theory of heat transfer. We have provided a detailed numerical and theoretical analysis of the singular resonance condition in moving media and predicted the existence of a giant nonequilibrium vacuum friction. Our earlier work discussed the non-local hydrodynamic model\cite{guo_singular_2013} and effect on the singular resonance while here we have shown that the result is valid in the relativistic limit.

A direct experimental investigation of our predicted effect is difficult due to constraints on the velocities to achieve this resonance however the delicate phase and amplitude balance (PAB) condition can emerge as a ubiquitous principle to be applied in moving nanophotonic media. Non-equilibrium opto-mechanical structures can also lead to light amplification effects which rely on our singular resonance condition.

\end{document}